# Noninvasive method for electrocardiogram recording in conscious rodents with the electro-conductive liquid electrodes

## © Valery Mukhin

Institute of Experimental Medicine, St. Petersburg, Russia

e-mail: Valery.Mukhin@gmail.com

*Existing methods of heart rate recording in animals have shortcomings, which significantly obscure the influence of experimental factors on heart rate. We have developed a method of electrocardiographic recording of heart rate in rats without these drawbacks. To contact the animal limbs used liquid electrodes which are two small baths filled with conductive fluid (saline solution). To provide the relative immobility (and quiet) of the animal the two baths was covered with a dark chamber without a bottom and with an entrance for the rat. During the experiment, a rat placed near the chamber comes into it (for the innate preference for darkness) and locates itself inside with its head for the exit. At that moment ECG recording starts. This method allows to record heart rate in the intact rodents (without anesthesia and stress) and does not require substantial preparation. It is not suitable for standard ECG analysis of the heart condition and function, but this is a good way for recording heart rate for the further analysis of its variability.*

Analysis of heart rate variability can not be considered as a valid tool for assessment of autonomic cardiovascular regulation due to complex and largely unexplored physiological basis underlying this phenomenon (Taylor and Studinger, 2006). Therefore, further study of underlying physiological mechanisms is actual and increasingly important (Fig. 1). However, such studying is not possible without animal experiments. To provide the relative immobility (and quiet) of the animal the two baths was covered with a dark chamber without a bottom and with an entrance for the rat.

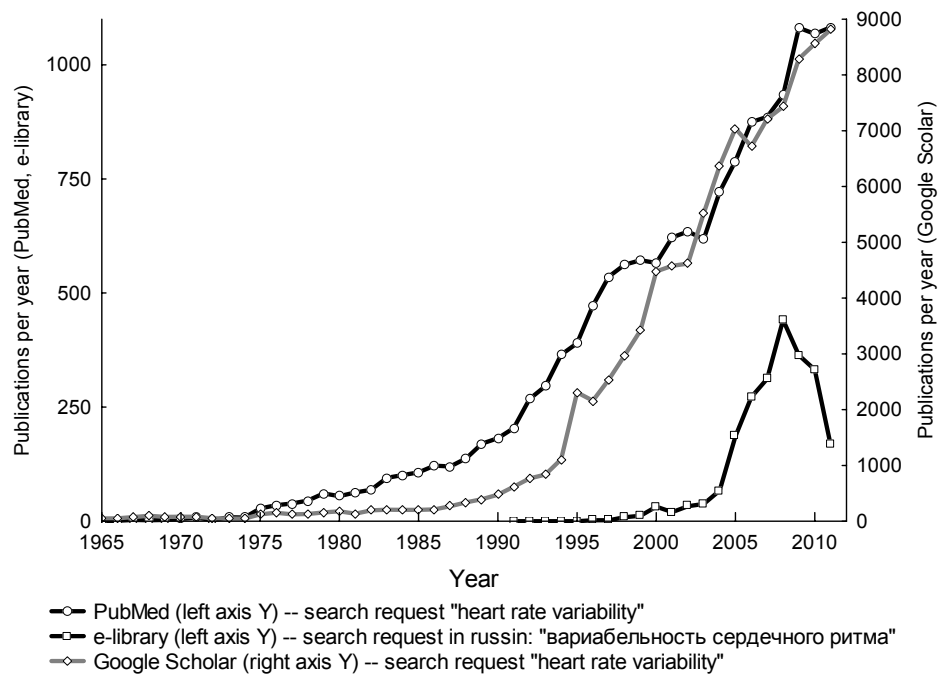

Figure 1. Publications per year on heart rate variability[*].

Existing methods of heart rate recording in animals have several shortcomings, which significantly obscure the influence of experimental factors. For example, some methods require anesthesia. Among them are surgery for ECG electrode implantation and recording of cardiac rhythm in anesthetized animals. But this is a bad way, as drug affects the heart rate both during operation (Kato et al., 1992; Latson et al., 1992; Deutschman et al., 1994; Sleigh and Donovan, 1999) and after anesthesia recovery (Latson et al., 1994). Another way to record ECG is to limit the freedom of the animal using specially designed boxes and jackets (Pereira-Junior et al., 2010). But confinement of animal leads to immobilization stress (Renaud, 1959) and therefore changes the heart rate variability. Finally all of the above methods require considerable time to prepare for ECG registration: handling, operation for electrode implantation, fixation of the electrode jacket, etc.

We have developed a method of electrocardiographic recording of heart rate in rats without these drawbacks. To contact the animal limbs used liquid electrodes which were two small baths (we used Petri dishes) filled with conductive liquid (saline solution) (Fig. 2). ECG recording in a liquid medium electrophysiologically justified and suggested as a way for ECG registration in the human completely immersed in the conductive fluid (Kinsht et al., 2009). Dimensions of the electrode baths was 10 * 10 cm, height was about 1 cm. Metal electrocardiograph electrodes was immersed in the liquid on the side of the baths. The forelegs of the rat situated in one of the baths, and the rear legs were placed in the other one.

To provide the relative immobility (and quiet) of the animal the two baths was covered with a dark chamber without a bottom and with an entrance for the rat. The dimensions of the chamber were 18 * 11 * 8 cm. Selection of the appropriate size was important. The small height of the chamber does not allow the rat to make vertical stands losing contact with the liquid. The small length and width does not allow the rat to avoid contact with the liquid or to place all the limbs in one of the baths. The entrance was a hole on the narrow side of the dark chamber. It was a semi-circle with the diameter of 6 cm.

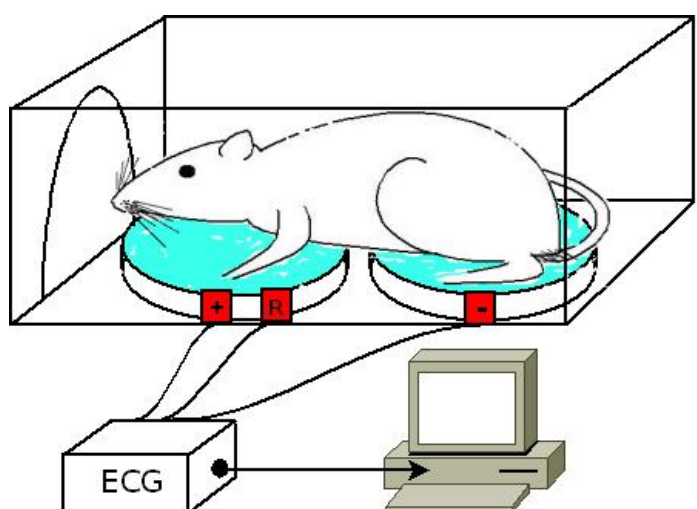

Figure 2. A rat in the experimental apparatus.

During the experiment, a rat placed near the chamber comes into it (for the innate preference for darkness), quickly explores it, turns around and locates itself inside with its head for the exit. At that moment ECG recording starts. ECG lead in this method is not standard, although it is close to the standard aVF one (Fig. 3). Once ECG registered, the dark camera gets up and the rat returns to the home cage.

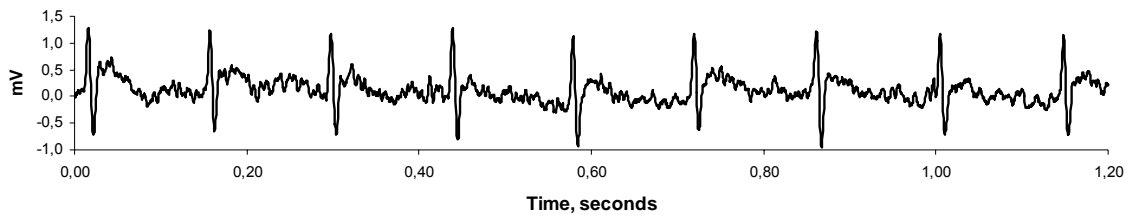

Figure 3. An electrocardiogram of the rat obtaining with the method.

The advantages of the method are the rat is conscious, relatively calm, the registration does not require prior manipulation of the animal and requires minimal amount of time (2-3 minutes per animal).

However, the method has two features that do not allow assessing of heart function itself. First, the point of the rat body which is in contact with the electrode is not constant. Therefore, the form of the electrocardiogram is not constant too. Another feature is the small current interference which is the result of induction in the liquid medium and the muscle activity of the animal. The interferences make it difficult to read small ECG waves. However, these features do not prevent detection of the big R-waves (heart rate) for the further analysis of series of them.

So, this method allows to record heart rate in the intact rodents (without anesthesia and stress) and does not require substantial preparation. It is not suitable for standard ECG analysis of the heart condition and function, but this is a good way for recording heart rate for the further analysis of its variability.